\documentclass[conference]{IEEEtran}
\IEEEoverridecommandlockouts
\usepackage{cite}
\usepackage{amsmath,amssymb,amsfonts}
\usepackage{graphicx}
\usepackage{textcomp}
\usepackage{xcolor}
\usepackage{comment} % Comentar texto com begin/end
\usepackage[ruled]{algorithm}
\usepackage[noend]{algpseudocode}
\usepackage{booktabs}
\usepackage{subcaption}
\usepackage{setspace}
\usepackage{multirow}

\algrenewtext{EndWhile}{\algorithmicend\ \algorithmicwhile}
\algrenewtext{EndFor}{\algorithmicend\ \algorithmicfor}
\algrenewtext{EndIf}{\algorithmicend\ \algorithmicif}
\algrenewtext{EndFunction}{\algorithmicend\ \algorithmicfunction}

\usepackage[normalem]{ulem} % tachar uma palavra

\def\BibTeX{{\rm B\kern-.05em{\sc i\kern-.025em b}\kern-.08em
    T\kern-.1667em\lower.7ex\hbox{E}\kern-.125emX}}
\begin{document}

% % Ajusta espaço entre as referências
% \newlength{\bibitemsep}\setlength{\bibitemsep}{.3\baselineskip plus .05\baselineskip minus .05\baselineskip}
% \newlength{\bibparskip}\setlength{\bibparskip}{0pt}
% \let\oldthebibliography\thebibliography
% \renewcommand\thebibliography[1]{%
%   \oldthebibliography{#1}%
%   \setlength{\parskip}{\bibitemsep}%
%   \setlength{\itemsep}{\bibparskip}%
% }
%\def\bibfont{\small}

\title{Social-based Cooperation of Vehicles for Data Dissemination of Critical Urban Events}

\begin{comment}

\author{\IEEEauthorblockN{1\textsuperscript{st} Alisson Yury}
\IEEEauthorblockA{\textit{Centre of Informatics} \\
\textit{Federal University of Para\'iba}\\
Jo\~ao Pessoa, Brazil \\
email address}
\and
\IEEEauthorblockN{2\textsuperscript{nd} Given Name Surname}
\IEEEauthorblockA{\textit{dept. name of organization (of Aff.)} \\
\textit{name of organization (of Aff.)}\\
City, Country \\
email address}
\and
\IEEEauthorblockN{3\textsuperscript{rd} Given Name Surname}
\IEEEauthorblockA{\textit{dept. name of organization (of Aff.)} \\
\textit{name of organization (of Aff.)}\\
City, Country \\
email address}
\and
\IEEEauthorblockN{4\textsuperscript{th} Given Name Surname}
\IEEEauthorblockA{\textit{dept. name of organization (of Aff.)} \\
\textit{name of organization (of Aff.)}\\
City, Country \\
email address}
\and
\IEEEauthorblockN{5\textsuperscript{th} Given Name Surname}
\IEEEauthorblockA{\textit{dept. name of organization (of Aff.)} \\
\textit{name of organization (of Aff.)}\\
City, Country \\
email address}
\and
\IEEEauthorblockN{6\textsuperscript{th} Given Name Surname}
\IEEEauthorblockA{\textit{dept. name of organization (of Aff.)} \\
\textit{name of organization (of Aff.)}\\
City, Country \\
email address}
}
\end{comment}

\author{
\IEEEauthorblockN{
{\bf Alisson Yury\IEEEauthorrefmark{1}},
{\bf Everaldo Andrade\IEEEauthorrefmark{1}},
{\bf Michele Nogueira\IEEEauthorrefmark{2}},
{\bf Aldri Santos\IEEEauthorrefmark{2}}, and
{\bf Fernando Matos\IEEEauthorrefmark{1}}
}
\IEEEauthorblockA{\IEEEauthorrefmark{1}Dept. of Computer Science - Federal University of Para\'iba, Brazil \\
\IEEEauthorrefmark{2}NR2 - Federal University of Paran\'a, Brazil, \\
Emails:  
alisson.yury@cc.ci.ufpb.br,
everaldo.andrade@ppgi.ci.ufpb.br, 
\{michele, aldri\}@inf.ufpr.br,
fernando@ci.ufpb.br}}

\maketitle

\begin{abstract}
Critical urban events need to be efficiently handled, for instance, through rapid notification. 
%Vehicular Ad Hoc Networks (VANETs)
VANETs are a promising choice in supporting notification of information on arbitrary critical events. Although the dynamicity of VANETs compromises the dissemination process, the connections among vehicles based on users' social interests allow for optimizing message exchange and data dissemination. This paper introduces SOCIABLE, a robust data dissemination system for critical urban events that operates in a SIoV network. It is based on vehicles' community with common interests and/or similar routines  and employs social influence of vehicles according to their network location to select relay vehicles. In a comparative analysis on NS3 with the MINUET system, SOCIABLE achieved 36.56\% less packets transmitted in a dense VANET and a maximum packet delivery delay of 28ms in a sparse VANET, delivering critical event data in a real-time and robust way without overloading the~network.
\end{abstract}

%\begin{comment}
%\end{comment}

\vspace{-0.3cm}
\section{Introduction}

%\fm{trocar a referência do Everaldo para o do PMC}

%\aly{Adicionar agradecimento à CAPES}

One of the main challenges on urban environments involves dealing with critical events that often arise randomly over the time and space. Critical events are commonly ones that, if not dealt with efficiently, can damage the smooth running of the city (e.g. road cracks, traffic jams) or endanger people's integrity (e.g. fires, accidents)~\cite{Monreal2018}.
% \begin{comment}
% \al{The ubiquitous infrastructures supported by Information and Communication Technologies - ICTs (communication networks, sensors, actuators, etc.), play an attractive role as facilitators in the detection and dissemination of urban events~\cite{andrade2020}.} 
% \end{comment}
Vehicular Ad Hoc Networks (VANETs) naturally appear as an option for supporting those essential tasks, 
% due to the ubiquity of vehicles and also, they do not suffer from computational resource limitations. 
as vehicles embed cameras and sensors capable of detecting events, and through Vehicle-to-Vehicle (V2V) and Vehicle-to-Infrastructure (V2I) communications, event data can be disseminated to external entities that handle the event. 
Meantime, the intermittent connections between vehicles due to 
%caused by 
the high topological dynamics of VANETs,~and the fact that urban area is not thoroughly covered by Base Stations (BSs) hinder data transmission, 
%Meantime, the high topological dynamics of VANETs
%and, thus, the intermittent connections makes it hard to transmit the data 
%\fm{V2I e visto que nem todas áreas são perfeitamente cobertas}
%\aly{through V2V, and also through V2I since urban area is not perfectly covered by BSs, (dificulta a transmissão de dados através de V2V e também através de V2I, uma vez que nem todas as áreas são perfeitamente cobertas por EBs)} 
so that solutions for dissemination depend almost exclusively on 
%\as{driven}
opportunistic connections~\cite{Vegni2015}. 
Moreover, the data traffic from  other applications running on VANET can damage the data flow dissemination of critical events~\cite{quadros2016qoe}. 
%Sapienza2016, 
Thus, vehicles may not be willing to cooperate to save their computational resources, exhibiting selfish behavior in the network~\cite{Rahim2018}. Those issues hamper the event treatment since their criticality demands a robust dissemination~\cite{Khan2015}.

%\al{A range of researches based on {\it Cyber Physical-Systems} (CPS) and {\it Cyber-Human-Systems} (CHS) have sought to employ new ways to build and coordinate systems in networks as a means of getting more intelligent autonomous management. Among these, the usage of social concepts has been gaining strength by allowing relationships between devices to be established from their users.} 
As part of the Cyber-Human-Systems (CHS) and Cyber Physical-Systems, the Social Internet of Things (SIoT) defines five types of object relationships: Parental (POR), relationship between devices from the same manufacturer; Co-work (C-WOR), relationship between objects that cooperate to meet an application’s demands; Co-location (C-LOR), relationships between objects~operating in the same location; Ownership (OOR), relationship between devices from the same user; and Social Objects Relationship (SOR), relationship between objects that come into contact sporadically or continuously. 
%The SIoT application in the VANETs’ environment gave rise to the 
The Social Internet of Vehicles (SIoV) rose as alternative for~messa\-ge sharing~between vehicles that hold a relationship. Usually, in SIoV, the relationships arise from the behavior of drivers, which can 
socialize to achieve a common goal~\cite{Silva2019}.~For instance, drivers passing through the same avenue at the same time are interested 
%in knowing about 
on road accidents that could~cause~delays.

The establishment and perception of community are two techniques of social networks employed in vehicular networks \cite{Wang2019}. %Although communities built in VANETs are not as strong as those in traditional social networks, 
We can use the temporal/spatial organization of communities to assist in the opportunistic dissemination of data from events that occur randomly in time/space. For instance, using social parameters to choose the relay nodes to forward the data can greatly influences in the dissemination performance \cite{xu2017}, by increasing the delivery rate, decreasing the average delivery delay \cite{ni2017} and reducing the propensity for selfish behavior. Social Network Analysis (SNA) techniques also assist in identifying influential components to improve the efficiency of network services \cite{Rahim2018b}. In the SIoV context, the network may have a small number of highly influential nodes, which regulate most of the connectivity and information flow. Such influential nodes may be vehicles that have connection advantages, thus impacting the dissemination of information.

This work proposes SOCIABLE (\textbf{SOC}ial mon\textbf{I}toring and dissemin\textbf{A}tion of ur\textbf{B}an critica\textbf{L} \textbf{E}vents), a dynamic system to carry out robust dissemination of data flow from critical urban event considering the social 
traits %characteristics 
of vehicles. SOCIABLE takes into account the 
%Social Object Relationships 
SOR established between vehicles 
with %that have 
common interests and/or routines to create temporary virtual communities. Communities and centrality techniques support a target dissemination by choosing relays to expedite data de\-li\-very, thus increasing user satisfaction (Quality of Experience – QoE) and improving performance and robustness of events’ dissemination (Quality of Service - QoS).
%\al{Simulation results in NS3 attest that SOCIABLE supports event data dissemination by collaboration between vehicles. Comparative tests against MINUET \cite{andrade2020}, a solution that does not employ social parameters, show that SOCIABLE come to similar or better results with less network overhead. }
Comparative tests in the NS3 simulator against MINUET
\cite{pmcandrade2020} 
%has %a solution that does not employ social parameters, 
%in the NS3 simulator, 
showed that SOCIABLE transmitted 36.56\% less packets and achieved~a~maximum packet delivery delay of 28ms,
indicating a better network~usage.
%In addition, the results also show that the social aspects influence in the delay and delivery of packets

The rest of the paper is organized as follows. Section~\ref{sec:rel} discusses the related work. Section~\ref{sec:prop} describes the SOCIABLE system and its operation. Section~\ref{sec:results} shows a performance evaluation and results achieved. Section \ref{sec:conc} concludes the paper.

\section{Related Work}
\label{sec:rel}

The ubiquity of vehicles %in cities 
has enabled  
%Due to the ubiquity of vehicles in cities, various solutions take into account
VANETs to assist in the daily lives of cities when dealing with a specific urban~aspect. However, most of %them 
solutions 
require that the vehicles 
actively participate and cooperate in the dissemination, which 
%usually 
does not take place in reality. For instance, devices with selfish behavior many times only cooperate in disseminating data with ones 
%those 
they have a strong social relationship or share a common interest in. Other~solutions rely on centralized entities or overload the network with duplicate messages.

In~\cite{Shrestha2018}, a centralized solution supports %efficient and reliable
dissemination of critical event messages on VANETs aided by a server in the cloud. Despite being able to carry out the dissemination, the vehicles need to send a request and receive the response from the server. The latency of communication with the cloud increases the response time, which can compromise efficiency, especially when considering critical events. %In~\cite{Yaqub2018} a scheme enables the dissemination of critical data in vehicular named data networks. Vehicles obtain data from multiple providers in the vicinity of the consumer vehicle using a single packet of interest. Although this scheme manages data forwarding efficiently, it does not deliver many packets in sparse networks and the cost of transmitting messages increases due to the temporary cache for each message on the network.
In \cite{pmcandrade2020}, a system called MINUET monitors and disseminates urban events based on a vehicular clustering approach. Although the vehicles cooperate to forward the data to a central authority, it applies a restricted flooding strategy, which overloads the network.

Other studies have advocated the benefits of social metrics to assist in the dissemination of data for real-time scenarios. In~\cite{Cunha2014},
vehicles exchange messages to find out their degrees and clustering coefficients to enhance message dissemination. Although the solution guarantees the delivery of messages, it imposes a high communication cost in low density scenarios due to message caching.
%by using clustering coefficient and node degree metrics, this solution finds nodes that behave as hub nodes, being they then good candidates to rebroadcast messages, since reach a high number of nodes, and thus a higher performance in data dissemination. Results show that the proposed solution guarantees the delivery of messages. However, although in high density environments it achieves a better performance when compared to other proposals, in low density environments it can increase the delivery delay. 
In~\cite{Campolo2018}, a social-enhanced framework facilitates data delivery with the support of social relationships of Vehicle-to-Everything (V2X) entities.
Although this solution obtained satisfactory results, it depends on centralized entities (switches and a server), which imposes limitations for real-time transmissions of critical events.

%\al{Although the authors prove  through a use case of disseminating alerts that the \textit{framework} can have better results when compared 
%using  the cellular network, the solution depends on devices on the edge (\textit{switches}) of the network and from a centralized server.}

The concern with selfish behaviors of network nodes has also been target in other works. In~\cite{Rahim2018b}, a socially-aware routing mechanism stimulates selfish nodes to cooperate in data forwarding. 
%This mechanism makes use of 
It uses social techniques
%(degree of centrality, knowledge of the local and global community, and social activity) 
and reputation to obtain the next relay node and ensure the forward of data. However, if a node with a better reputation than the current relay is not found, it stores the data until it can find a node with a better reputation, which increases the cost of transmitting messages. In~\cite{Wang2018}, a traffic management system based on crowdsensing encourages vehicles, by the means of virtual monetary incentive, to cooperate in message forwarding of abnormal events. It employs a hybrid strategy (mobile network and RSUs) for a better cost-benefit ratio. However, vehicles can help spread and not be rewarded, thus wasting resources.

%\section{Social monitoring and dissemination of urban critical events}
\section{The SOCIABLE System}
\label{sec:prop}
This section describes the SOCIABLE (\textbf{SOC}ial mon\textbf{I}toring and dissemin\textbf{A}tion of ur\textbf{B}an critica\textbf{L} \textbf{E}vents) system for collaborative monitoring of 
%\nota{precisa redefinir as siglas aqui?} 
critical urban events (CUE) based on SIoV (Social Internet of Vehicles). Initially we present the characteristics and infrastructure of the urban environment where SOCIABLE operates, as well as, its architecture and components. Then, we illustrate the relay selection and the SOCIABLE operation to disseminate a critical urban event.
%Our SIoV network assumes the existence of low delay, overload and package collisions. For this, each vehicle must carry out community recognition in order to identify the most appropriate vehicle, among the nodes of its community, to continue the dissemination process by considering social metrics, such as, degree of centrality.

%\vspace{-0.1cm}
\subsection{SIoV Environment}
\begin{figure}[ht]
    \centering
    \includegraphics[scale=0.4]{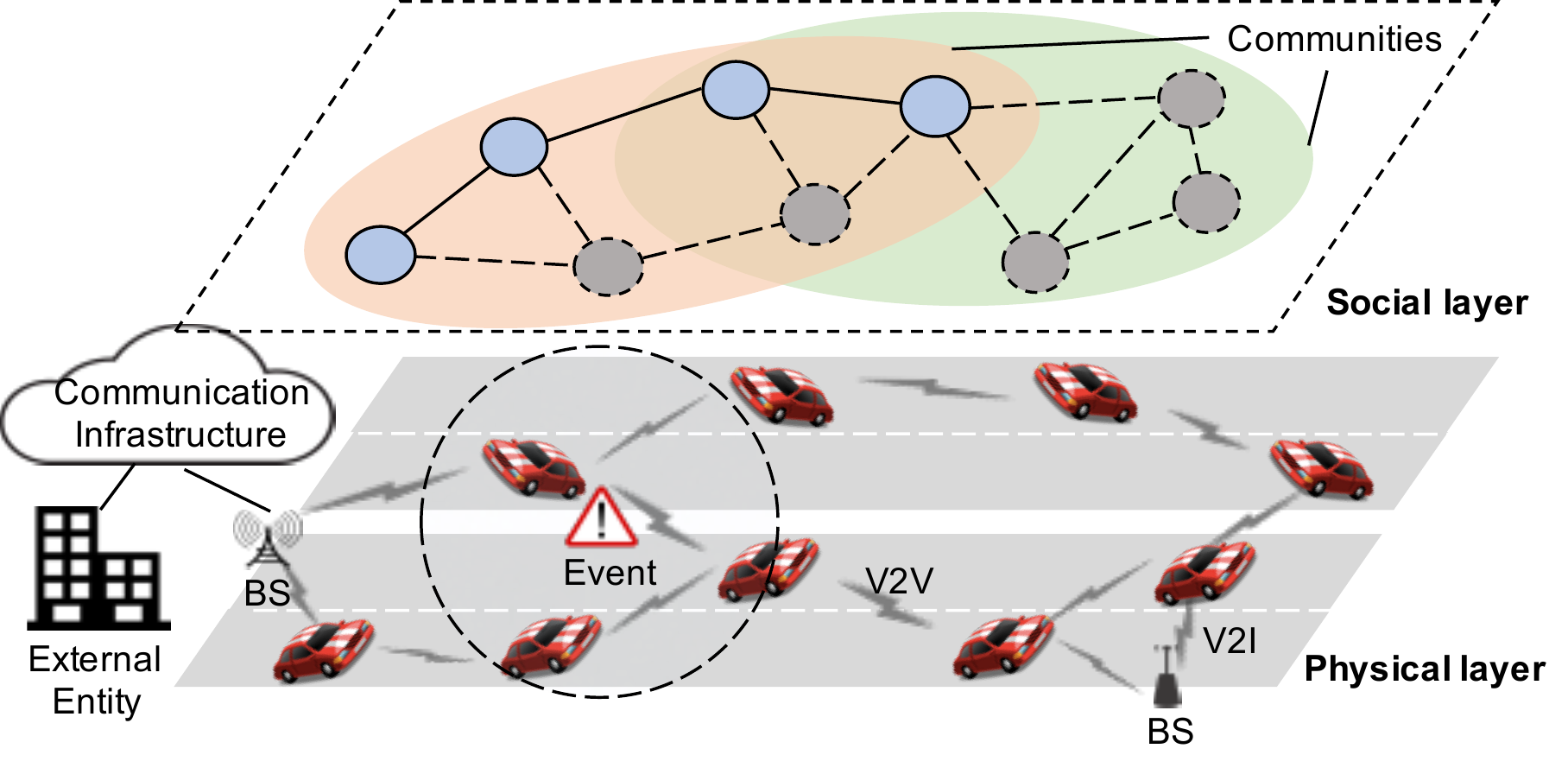}
    \caption{The SIoV model} %\as{Essa figura precisará ser melhorada para ficar mais compacta e também numa proporção melhor}
    \vspace{-0.3cm}
    \label{fig:model}
\end{figure}

SOCIABLE runs on an SIoV environment composed of two layers, as shown in Figure~\ref{fig:model}. The \textbf{Physical Layer (PL)} comprises vehicles that continuously travel on road networks and with different speeds. All vehicles communicate with other vehicles, Vehicles-to-Vehicles (V2V) or with Urban Infrastructure through Base Stations (BSs), Vehicles-to-Infrastructure (V2I), creating hybrid vehicular networks. CUEs can randomly arise in time and space. A CUE comprehends any event that impact the day-to-day life of the city and its citizens, such as fires, accidents, crimes, road obstructions, among others. These events must be usually notified to competent authorities, that mean External Entities (EE) responsible for handling events.

The \textbf{Social Layer (SL)} consists of the social profiles  of the vehicles built based on the social properties 
%\aly{(ex: rotina similar, interesse, local de trabalho)} 
of the drivers
(e.g. similar routines). Like any social network, nodes (vehicles) establish relationships that can be created and broken down over time. In addition, nodes with common interests or properties can be part of a community. For instance, nodes that travel on the same routes at the same  schedules (same routine) are able to be part of a community to exchange messages. Any node can attend in more than one community and nodes that are~part of the same community have a willingness to collaborate.~Thus, some (or all) nodes in a community cooperate to perform tasks as the dissemination of data from a CUE (blue nodes in Fig.~\ref{fig:model}).
% In this SIoV environment, vehicles that detect an event can initiate cooperation with other vehicles to disseminate information on the event, as long as they are part of the same community.

We denote an SIoV composed of a set of vehicles $V = \{v_ {1}, v_ {2},\dots, v_ {n}\}$ and a set of events $ E  = \{ev_ {1}, ev_ {2},\ldots, ev_ {m}\}$, so that any $v_i \in V$ vehicle can detect a $ev_j \in E$ event, as long as $ev_j$ is within the range of $v_i$. 
Vehicles that come into contact with each other hold Social Object Relationships (SOR), thus sharing a common profile and establishing communities.
%Vehicles that hold Social Object Relationships (SOR) by coming into contact establishes with each other
%\aly{estabelecidos a partir da frequência com que veículos entram em contato de forma esporádica ou contínua, os veículos podem formar comunidades com outros veículos que compartilhem o mesmo interesse/perfil social.}
%thus sharing the same social profile, establish communities. 
%In this way, 
The SIoV also makes a set of communities $C = \{c_1, c_2,\ldots, c_p \}$ where for any community $c_k \in C$, $c_k \subseteq V$. Therefore, $v_i \in c_k$ if, and only if, $v_i$ has the same interest/social profile as the~$c_k$ community. When a $v_i \in c_k$ vehicle detects a $ev_j$ event, the formation of a set $D(ev_j) \subseteq c_k$ is initiated, defined as a subset of cooperating vehicles from 
%the 
$c_k$ 
%community 
in the $ev_j$~spread.

\subsection{Architecture} 

\begin{figure}[ht]
    \centering
    \includegraphics[scale=0.35]{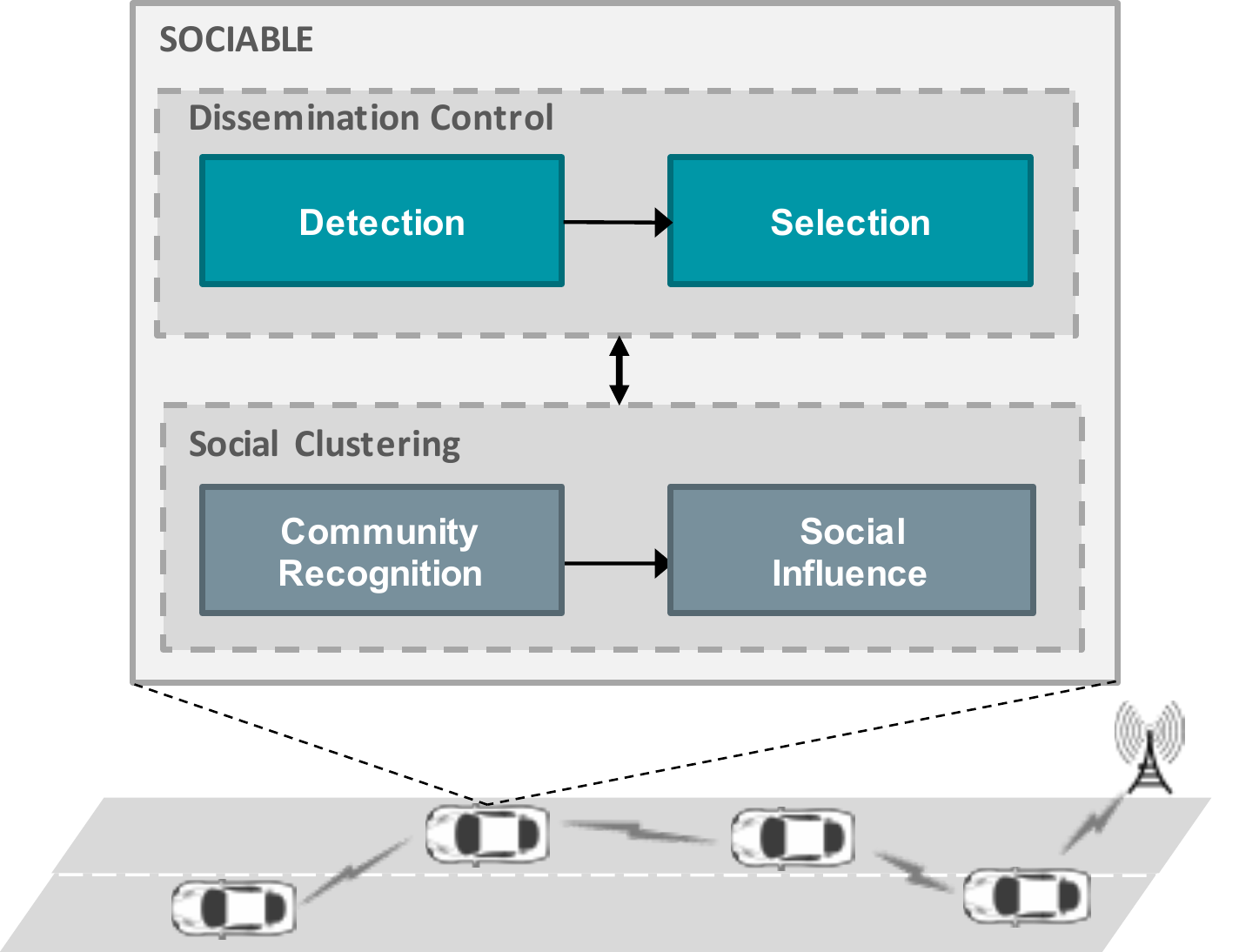}
    \caption{The SOCIABLE architecture}
    \vspace{-0.2cm}
    \label{fig:architecture}
\end{figure}

The SOCIABLE architecture holds two modules with two components each, as shown in Figure~\ref{fig:architecture}. The \textbf{Social Clustering Module} handles the management of social communities formed by vehicles that share a SOR.
%the Relationships (SOR),
%\as{the Social Object Relationships (SOR)}. 
The \textbf{Dissemination Control Module} coordinates the detection of the events and the selection of the vehicles to cooperate in the dissemination of the data of such events to the EEs.

The Social Clustering Module aims to create and maintain the social communities, and to  calculate the social significance 
%importance/significance/magnitude  
of the vehicle in its  community.
%that the vehicle has in the community. 
It consists of two components: \textbf{Community Recognition} and \textbf{Social Influence}. The \textbf{Community Recognition} component manages the SOR between vehicles by identifying the local community of the current vehicle. The local community are vehicles that are part of the same community as the current vehicle and are directly connected to it (neighbors). This is important because only vehicles belonging to the local community can cooperate to disseminate the event. Every $v \in V$ periodically transmits beacons containing its identification ($id$), current position ($pos$), number of neighbors ($nNbr$), and social profile ($profile(v)$) to build community knowledge. When a vehicle $v_i$ receives a beacon from a vehicle $v_j$, it checks whether $profile(v_i) == profile (v_j)$. If so, it stores $v_j$ in its list of neighbors willing to cooperate ($nbrList(v_i)$). The beacons are constantly exchanged to assist the community maintenance.

The \textbf{Social Influence} component deals with the social importance the current node has in the community%
, hence to its neighborhood. 
By means of the $nbrList$ and social metrics, this component calculates which neighbors are more likely to cooperate based on their social parameters. In this work, we employ the \textit{Structural Influence} (\textit{STR}) metric, 
which denotes how influential a node is based on its current position. As more influential a node is, higher its contribution on the data dissemination efficiency
%\aly{que denota o quão um nó é influente na rede baseado na sua posição atual, sendo que quanto mais influente é um nó, maior a sua contribuição na eficácia da disseminação de dados~}
\cite{Nagaraj2019}.
%which means how influential a node is based on its current position~\cite{Nagaraj2019}.

The \textbf{Dissemination Control Module} addresses the detection of the CUE, the selection of the cooperative vehicles and the forwarding of the event data flow to the EE.
%\as{events data} \tachar{to} \as{forward} the EE. 
This module consists of the \textbf{Detection} and \textbf{Selection} components. The first one detects and starts the monitoring of the events. In this way, when a given $vd$ vehicle is inside range of a $ev$ event, 
%\as{for instance,} 
$vd$ automatically assumes the role of \textit {Detector} and collects the context data of $ev$, which are: instant detection time, location, speed and direction. For the sake of simplicity, we assume that the analysis and detection of the CUE %\tachar{are carried out} 
take place through image processing and analysis techniques obtained by sensors from cameras embedded in the vehicles. We also assume that during the event analysis, the Detection component estimates the \textit{Time-to-Live} ($TTL$), which is the maximum number of hops over a message can be propagated.

When obtaining the context information of the event, the \textbf{Selection} component starts the process of defining the other two roles that nodes can play during the cooperation for the dissemination, which are: \textit{(i) Relay:} the vehicle responsible for forwarding the event data flow in the neighborhood, in addition to selecting the next relay node; \textit{(ii) Gateway:} vehicle that delivers the data about the monitored event to BS and, consequently, to EE. Considering the current node as the $vd$ detector node, the $c_k$ community and that $vd \in c_k$, $vd$ assumes as possible collaborators the nodes that are part of $c_k$. Thus, $\forall v_i \in c_k$, if $v_i$ is within reach of a BS, then $v_i$ is \textit{gateway}. If $v_i$ has the highest \textit{STR} among all neighbors of $vd$ in $c_k$, then $v_i$ is \textit{relay}. It is important to note that 
both the $vd$, after detecting an event, or any $v_i$ that has been selected as relay, can initiate the selection process.
%the selection process can be initiated by $vd$ after detecting an event, or by any $v_i$ that has been selected as relay. 
Once the relay is selected, $vd$ (or $vi$) generates monitoring packets and broadcasts them in the neighborhood. It also informs the $id$ of the next selected relay node. In addition, when $vd$ (or $vi$) is within reach of a BS, it delivers the event data to the BS.

\begin{algorithm}[ht] {
%\footnotesize
\setstretch{0.85}
\fontsize{9}{12}
\caption{Detection and dissemination}

\label{alg:control}
\begin{algorithmic}[1]
\Require{$msg.Mon(ev)$}
\Ensure{\textit{nbrList(v)} is the list of neighbors of \textit{v} in the community */}
\If{$ev$ is in $v_i$'s range}
\State $vd \gets v_i$
\State $vd$ creates $msg.Mon(ev)$;
\EndIf

\If{$TTL > 0$}
\If{$v_i$ is within a $BS$'s range}
\State $v_i$ is \textit{gateway}
\State $v_i$ delivers $msg.Mon(ev)$ to $BS$
\EndIf

\If{($v_i == vd$) or ($v_i == vr$)}
\For{each vehicle $v_j$ in $nbrList(v_i)$}
\State check $STR$ of $v_j$
\EndFor
\State $vr \gets v_j$ with the highest $STR$
\State $v_i$ disseminates $msg.Mon(ev)$ in the neighborhood
\EndIf
\Else
\State discards $msg.Mon(ev)$
\EndIf
\end{algorithmic}
} % fim do tamanho
\end{algorithm}

Algorithm~\ref{alg:control} describes the SOCIABLE operation when a $v_i$ vehicle detects a $ev$ event or when $v_i$ receives a $msg.Mon(ev)$ monitoring message. When $v_i$ detects $ev$, it starts to act as the $vd$ detector vehicle and creates $msg.Mon(ev)$ (lines 1--4). Regardless of whether $v_i$ received or created $msg.Mon(ev)$, $v_i$ checks that its $TTL$ has not yet expired (line 5). Case the $TTL$ has expired, $v_i$ discards the message. Otherwise, $v_i$ checks whether it is within range of an BS. If so, then $v_i$ plays as \textit{gateway}, thus delivering the message (lines 6--9). If $v_i$ is $vd$ detector or $vr$ relay (line 10), then $v_i$ initiates community recognition. Upon recognition, $v_i$ checks which of its neighbors has the highest \textit{STR}. This neighbor is then informed it is the next $vr$ relay. Finally, $v_i$ disseminates $msg.Mon(ev)$ (lines 11--17).

\subsection{Relay selection}
\label{sub:relay}
SOCIABLE employs a strategy to achieve QoS and QoE gains when selecting the relays. QoE values are based on social benefits the devices offer, which are perceived by employing SOR. SOCIABLE chooses as the next relay the neighbor node with the highest \textit{STR}, which is calculated according to Equation~\ref{eq:ac} \cite{Nagaraj2019}. It makes use of four centrality metrics:
%\cite{freeman1978, bonacich1987}
(i) \textit{Betweenness Centrality} (\textit{BC}), corresponds to how many times a node is on the shortest path between all pairs of nodes; (ii) \textit{Closeness Centrality} (\textit{CC}), denotes how close a node is to all other nodes; (iii) \textit{Degree Centrality} (\textit{DC}), is the number of nodes directly connected to a node (i.e. its neighbors); (iv) \textit{Eigenvector Centrality} (\textit{EC}), corresponds to how many important nodes a node is directly connected to. In this work, we assume
%major/meaningful/significant/magnitude/substantial, 
important nodes as those that directly reach BSs (i.e. gateways). Each metric has an associated weight (\textit{W}) and the sum of the four weights is 1.
\vspace{-0.1cm}
\begin{equation}
\footnotesize
\begin{split}
STR_i = W_{BC} * (BC)_i +W_{CC} * (CC)_i \\
+ W_{DC} * (DC)_i + W_{EC} * (EC)_i
\end{split}
\label{eq:ac}
\end{equation}

\begin{figure}[ht]
    \centering
    \includegraphics[scale=0.3]{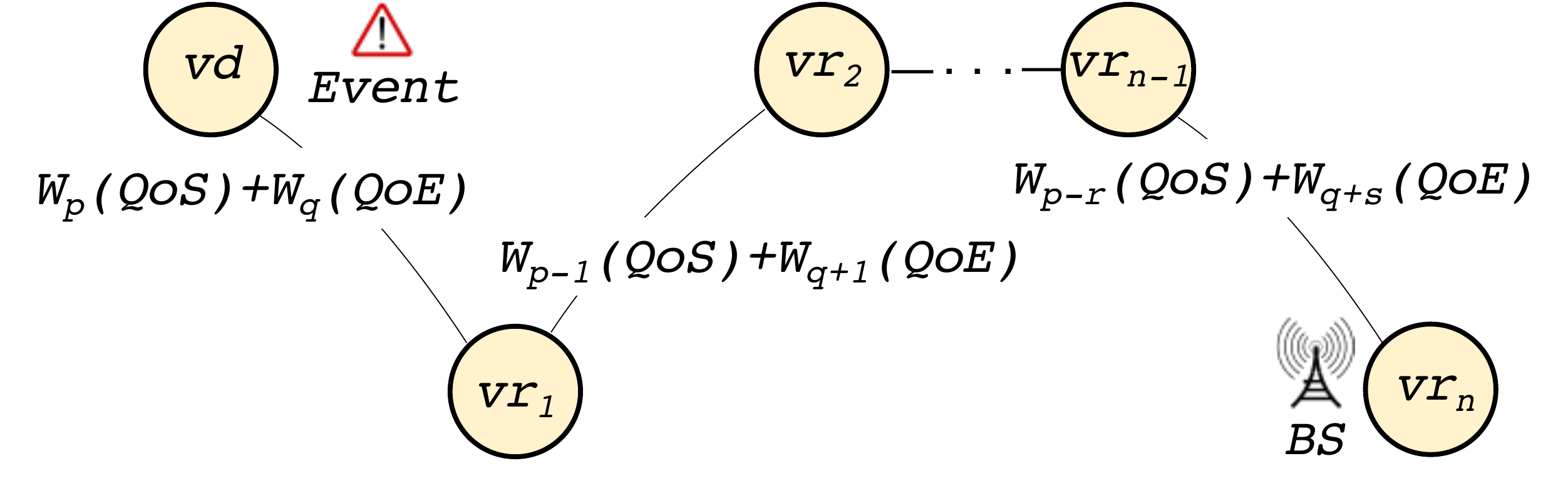}
    \caption{Relay selection}
    \vspace{-0.3cm}
    \label{fig:relay_selection}
    \vspace{-0.2cm}
\end{figure}

The first three metrics usually improve QoS, since choosing a node with better \textit{BC}, \textit{CC} and \textit{DC} values increases the probability in delivering the data \cite{Rahim2018}, while the \textit{EC} metric influences the QoE since if the data reaches the gateway quickly, the user satisfaction increases. Thus, when a detector vehicle selects the next relay to forward the packet, it gives more importance to QoS (higher $W_{BC}$, $W_{CC}$ and $W_{DC}$ values) and less to QoE (lower $W_{EC}$ value). As the next relays are chosen, the influence of QoS decreases while QoE increases. Figure \ref{fig:relay_selection} illustrates the relay selection process. When \textit{vd} detects a event, \textit{vd} chooses the next relay by applying a higher weight to QoS than QoE ($w_p > w_q$). As the next relays are chosen, $w_p$ value decreases while $w_q$ value increases. This process continues until the packet reaches a gateway.

\subsection{Operation}

\begin{figure}[ht]
    \centering
    \includegraphics[scale=0.275]{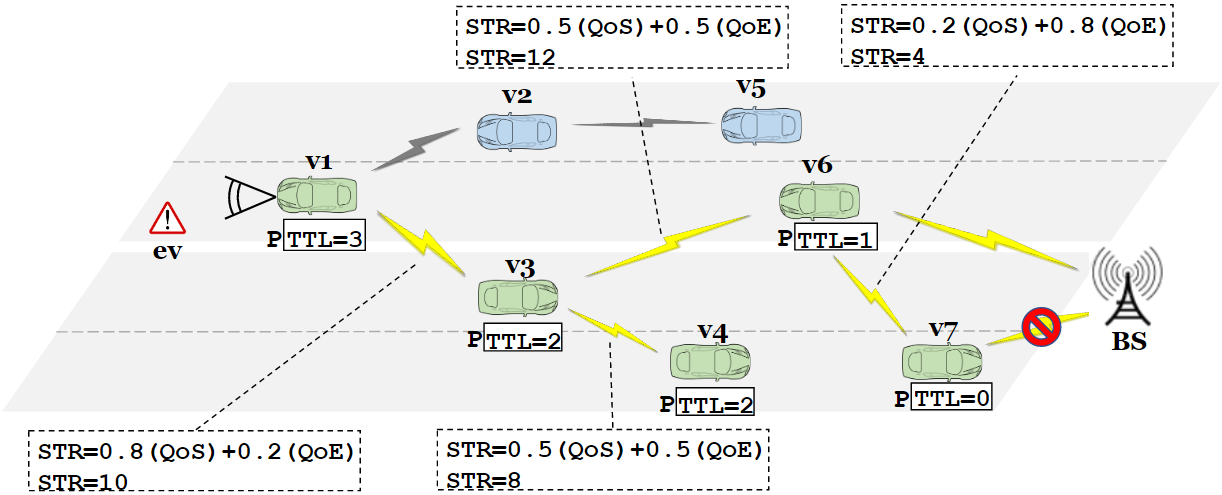}
    \caption{Example of the SOCIABLE operation}        \vspace{-0.3cm}
    \label{fig:operation}
\end{figure}

Figure~\ref{fig:operation} illustrates SOCIABLE's operation when detecting an $ev$ event. In the example scenario, we have the set $V = \{v1, v2, v3, v4, v5, v6, v7\}$ and a BS to assist in the $ev$ data transmission. We have also established two communities $C1$ and $C2$ formed by green and blue vehicles, respectively. In this way, when $v1$ detects $ev$, it simultaneously starts the community recognition and monitoring step, thus it creates a $P$ packet with a $TTL = 3$ value, which limits the number of transmissions of $P$. During community recognition, $v1$ finds that, among its neighbors, only $v3$ is part of its $C1$ community. Thus, regardless its \textit{STR} value, $v3$ is chosen as the next relay. Whereas with each $P$ transmission, its $TTL$ decreases by one unit, so that when $v3$ receives $P$, its $TTL = 2$, which means that $v3$ can then retransmit the $P$ packet.

When relaying $P$, $v3$ discovers that it has two neighbors, $v4$ and $v6$, in $C1$. Among them, $v3$ chooses the one with the highest \textit{STR}, thus being able to reach more nodes in the network. In this case, we consider $v6$ (\textit{STR} $=12$) as the node with the highest \textit{STR}, thus being chosen as the relay. $v3$ then transfers $P$ to $v4$ and $v6$. As $P$ is retransmitted, the relay increases the weight (influence) of the QoE indicator while decreasing the weight of QoS indicator. Since $v4$ is not a relay, nor is within reach of a BS, it discards $P$. On the other hand, $v6$ is within reach of a BS (\textit{gateway}), then it delivers $P$. In addition, since $v6$ is also a relay, it passes $P$ to $v7$. Finally, although $v7$ is also within reach of a BS, it finds that the $TTL$ of $P$ has expired ($TTL = 0$), thus discarding $P$. In this example scenario, as the relay nodes are chosen, the group $D(ev)= \{v1, v3, v6, v7\}$ is established, which are the vehicles that cooperated to disseminate the event data.

\section{Evaluation}
\label{sec:results}

We evaluate SOCIABLE to assess its efficiency and robustness in monitoring and disseminating the data flow of critical urban events, as well as, the social aspects of the system and their influence and benefits to the dissemination. We have implemented and tested in the NS3 simulator, version 3.28, together with SUMO. We also made use of the LuST project (Luxembourg SUMO Traffic), a real traffic mobility scenario~\footnote{https://github.com/lcodeca/LuSTScenario}. The simulations were carried out in two different scenarios, considering one event, different numbers of traffic lanes, direction of the road and number of vehicles traveling.
%\aly{utilizando dados de mobilidade estáticos que simulam tráfego realístico}
We set up a single fixed event with a total duration of 8 minutes. We take into account communities formed by vehicles that realize the same routine. That is, being a vehicle $v$, a time interval $\Delta t$ and an area $a$, we have that $v(\Delta t,a)$ denotes that the vehicle $v$ is in the area $a$ in the interval $\Delta t$. Thus, two vehicles $v_i$ and $v_j$ establishes a SOR relationship when $v_i(\Delta t_i,a_i) == v_j(\Delta t_j,a_j)$. Therefore, we use a vehicle relationship rate of 90\%. That is, 90\% of the vehicles in the test scenarios are part of the community. Despite being a high rate, it can be achieved if we consider drivers who travel on a stretch of road at the same time, for example, when taking their children to school. We compared the performance of SOCIABLE with MINUET, which is a solution that does not contemplate social characteristics.

%Figure~\ref{fig:cutout} exhibits two cutouts from LuST meaning high density (HD) and low density (LD) scenarios, and also routes used in the simulation. The yellow circles correspond to the positions where events on the roads occur. 
\vspace{-0.1cm}
\begin{table}[H]
\centering
\caption{Simulation setting}
\label{tab:parameters}
\footnotesize
\begin{tabular}{lll} \hline
\toprule 
\textbf{Parameter} & \textbf{LD} & \textbf{HD}\\ 
\midrule        
Simulation time & 9 mins & 9 mins\\
%Number of events & 1 & 1\\
Event duration & 8 mins & 8 mins\\
Kind and number of lanes & one-way / 3 lanes & two-way / 6 lanes\\
%Number of lanes & 3 & 6\\
Transmission range & 100 m & 100 m\\
Number of vehicles &91 & 754\\
Number of base stations &1 & 1\\
\bottomrule 
\end{tabular}
\end{table}
\vspace{-0.2cm}
Table~\ref{tab:parameters} shows the parameters applied in the simulations for low density (LD) and high density (HD) scenarios. We assess the SOCIABLE efficiency through the following metrics: Number of Collaborating Vehicles ({\bf NCV}), which indicates the amount of vehicles (detector, relay and gateway) that collaborated to the dissemination; Average Delivery Delay ({\bf ADD}), which denotes to the time interval between the instant the event is detected and the first monitoring packet is delivered to the BS; Number of Generated Monitoring packets ({\bf NGM}), which indicates the total amount of monitoring packets generated by the detecting vehicles; Number of Delivered Monitoring packets ({\bf NDM}), which corresponds to the total amount of monitoring packets delivered to the BS; Exchanged Packet Overload ({\bf EPO}), that means the communication cost (beacons and monitoring packets exchanged) by running SOCIABLE. To evaluate the SOCIABLE social behaviour, we assessed the STR metric, which encompasses four other social metrics, as explained in Subsection \ref{sub:relay}.
%We increased the QoS social metric weights from 0,1 to 0,9 and decreased the QoE social metric weight from 0,9 to 0,1 as the next relays were chosen.
Over the relays' selection, we increased the QoS social metric weights from 0,1 to 0,9 and decreased the QoE social metric weight from 0,9 to 0,1.

%\aly{com pesos: .1, .25, .33, .4, .5, .6, .66, .75 e .9. As métricas de centralidade para QoS começam em .1 e a cada salto aumenta um peso, com a métrica de centralidade para QoE ocorre o inverso, começa em .9 e a cada salto diminui um peso}.

\begin{comment}
\begin{figure}[ht]
	\centering
		\includegraphics[width=65mm]{Images/scenarios.pdf}
		\caption{LuST cutouts}
		\label{fig:cutout}
\end{figure}

\end{comment}

\subsection{Results}

\begin{figure}[!htb]
\vspace{-0.3cm}	
	\includegraphics[scale=0.21]{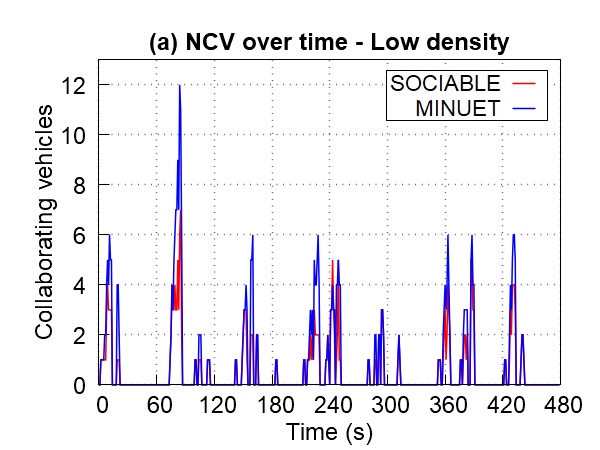}
	\hspace{-0.5cm}
	\includegraphics[scale=0.21]{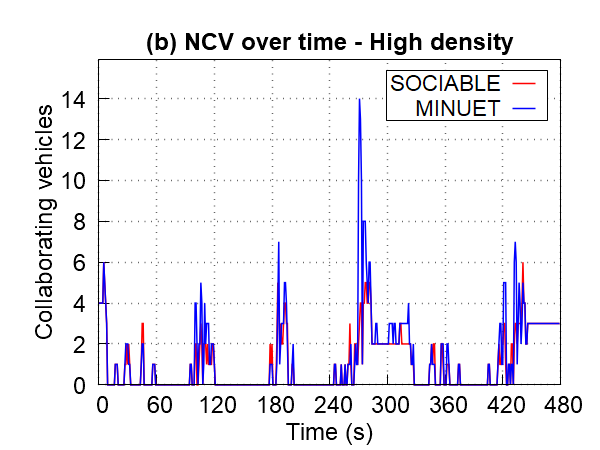}
	\caption{Vehicles collaborating in the dissemination}
	\vspace{-0.3cm}
	\label{fig:nvcXt-MINUET}

\end{figure}

This section presents the analysis of the tests carried out to evaluate the performance of SOCIABLE compared to MINUET, in addition to SOCIABLE's social behavior. The graphs in Figure \ref{fig:nvcXt-MINUET} show the number of vehicles collaborating to disseminate the event data over time (NVC) in the high density (HD) and low density (LD) scenarios. Unlike MINUET, which uses a restricted flooding, the targeted dissemination strategy of SOCIABLE results in fewer collaborating vehicles (a smaller NCV value) in both scenarios. In the LD scenario 
%(Graph \ref{fig:nvcXt-MINUET}(a))
up to 7 and 12 vehicles collaborated in SOCIABLE and MINUET, respectively, while in the HD scenario,  
%(Graph \ref{fig:nvcXt-MINUET}(b)), 
a maximum of 6 vehicles collaborated in SOCIABLE and 14 in MINUET. A greater NCV value can indicate overuse of resources, since similar or even better results were achieved with fewer collaborating vehicles.
%\al{In addition, by limiting dissemination to members of the community only, SOCIABLE prevents selfish behavior~in~vehicles.} \as{Como não testamos isso então podemos afirmar. Assim,melhor remover, isto  é, comentar o texto}. 

\begin{figure}[!htb]
	\includegraphics[scale=0.21]{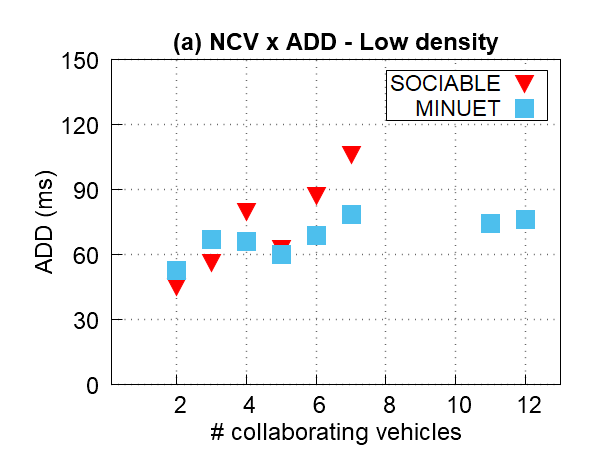}
	\hspace{-0.5cm}
	\includegraphics[scale=0.21]{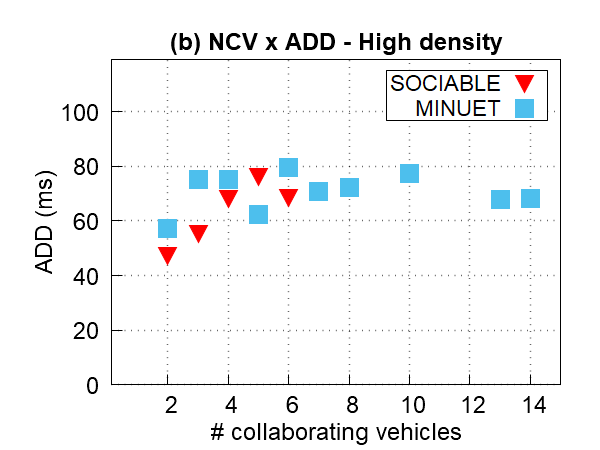}
	\caption{Average delivery delay of monitoring packets}
	\vspace{-0.1cm}
	\label{fig:nvcXameMINUET}
\end{figure}

The graphs in Figure \ref{fig:nvcXameMINUET} show the average delivery delay (ADD) of generated monitoring packets over the number of vehicles that contributed to deliver the packets. In both scenarios, LD and HD, 
%(Graphics \ref{fig:nvcXameMINUET}(a) and \ref{fig:nvcXameMINUET}(b),respectively), 
fewer vehicles collaborated in SOCIABLE to deliver the packets. In LD, there is an upward trend of the ADD of both SOCIABLE and MINUET of up to 7 vehicles, where SOCIABLE delivered a maximum delay 35.26\% higher than that of MINUET.
%, \aly{porém, SOCIABLE entrega pacotes com atraso médio de 56ms enquanto MINUET entrega em 60ms.} %which is still an adequate delay for real-time transmissions. 
In HD, both systems had similar ADD values until 6 vehicles collaborating. Beyond that, SOCIABLE did not deliver monitoring packets, while MINUET's ADD stabilized.
%\aly{Neste cenário, SOCIABLE possui atraso médio de 54ms e MINUET possui atraso médio de 61ms.} 
Nonetheless, those results demonstrate that SOCIABLE delivers packets in real time and achieves higher efficiency in resource utilization (fewer number of vehicles to deliver packets).  Thus, it enables good user satisfaction (QoE) while achieves good QoS. In addition, the overall ADD of SOCIABLE is lower than the MINUET´s. In  LD, the ADD values of SOCIABLE and MINUET were 56ms and 60ms, respectively, while in HD the ADD values were 54ms and 61ms, respectively.

\vspace{-0.3cm}

\begin{figure}[!htb]
	\includegraphics[scale=0.21]{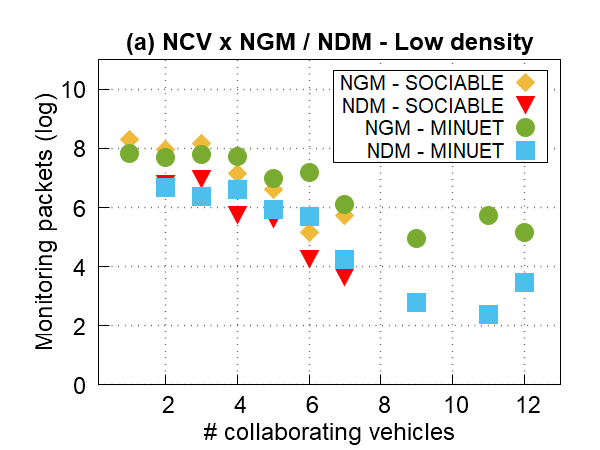}
	\hspace{-0.5cm}
	\includegraphics[scale=0.21]{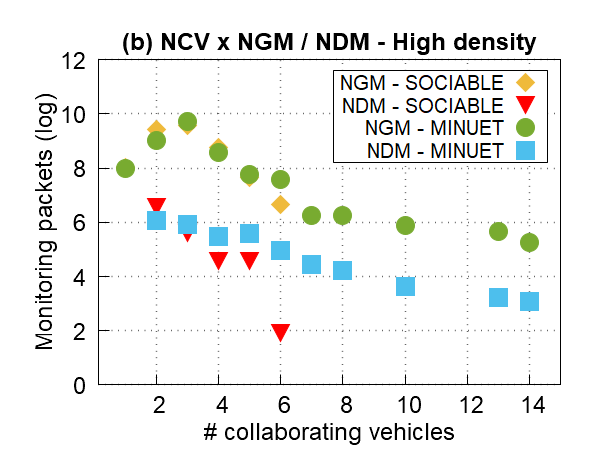}
	\caption{Average delivery delay of monitoring packets}
	\vspace{-0.1cm}
	\label{fig:nvcXnpmcXnpme}
\end{figure}

The graphs in Figure \ref{fig:nvcXnpmcXnpme} show the number of monitoring packets created (NGM) when the event is detected and monitoring packets delivered (NDM) to the BS, over the number of collaborating vehicles (NCV) in SOCIABLE and MINUET. In both scenarios,  
%(Graphics \ref{fig:nvcXnpmcXnpme}(a) and (b)), 
SOCIABLE and MINUET obtained same NGM values, despite fewer vehicles collaborated in SOCIABLE. This is because there are the same number of detecting vehicles. In LD scenario, 
%\aly{SOCIABLE possui taxa de entrega de 20\% e MINUET possui taxa de entrega de 22\%.} 
SOCIABLE achieved a NDM value higher than MINUET when there are fewer vehicles collaborating. As more vehicles collaborate, MINUET delivers more packets than SOCIABLE. In HD scenario,
%\aly{SOCIABLE entrega 3\% dos pacotes criados enquanto MINUET entrega 4,3\% dos pacotes criados.} 
SOCIABLE achieved a higher NDM than MINUET only when two vehicles collaborate. Furthermore, 
%In this scenario, 
MINUET delivered packets with up to 14 collaborating vehicles. The number of vehicles collaborating is higher in MINUET because it uses flooding, whereas SOCIABLE uses social parameters to choose one relay only, which reduces the number of copies of packets traveling on the network. Results indicate that SOCIABLE and MINUET are able to detect and deliver monitoring packets of urban events to the BS under different network and traffic conditions. The results also attest the higher efficiency in resource utilization of SOCIABLE, since fewer vehicles collaborate to deliver similar amount of monitoring packets.

\vspace{-0.3cm}

\begin{figure}[!htb]
	\includegraphics[scale=0.21]{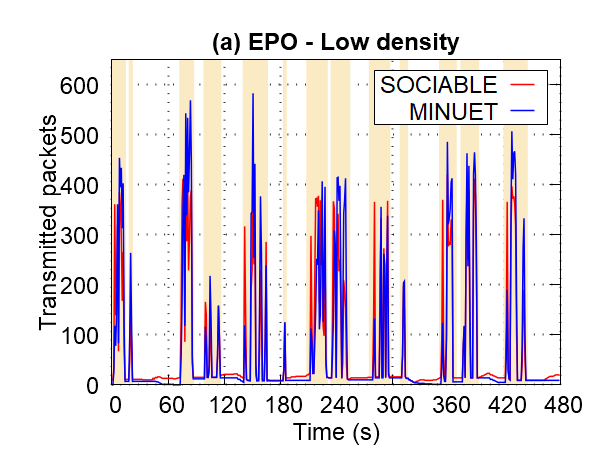}
	\hspace{-0.5cm}
	\includegraphics[scale=0.21]{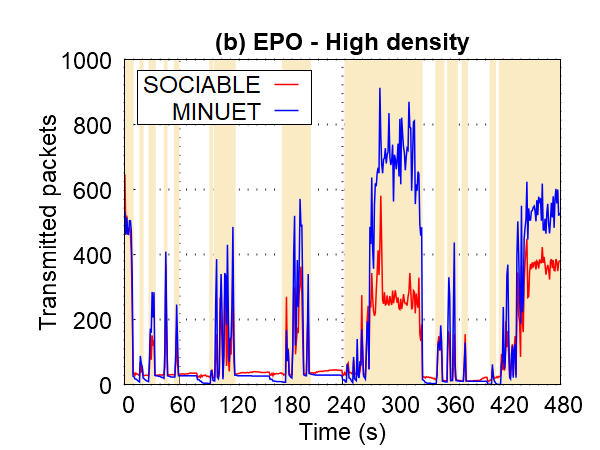}
	\caption{Communication cost under low and high densities}
    \vspace{-0.1cm}
	\label{fig:nptrXt-compare}
\end{figure}

The graphs in Figure \ref{fig:nptrXt-compare} show the exchanged packet overload (EPO) of SOCIABLE and MINUET in the LD and HD scenarios. Due to the employed targeted dissemination, SOCIABLE transmitted 3.49\% and 36.56\% less packets than MINUET in  LD and HD scenarios. Thus, together with the previous results, demonstrate how SOCIABLE's social parameters enable a better network resource usage while being able to carry out the monitoring, dissemination and delivery of packets.

\vspace{-0.3cm}

\begin{figure}[!htb]
%\hspace{-0.2cm}
	\includegraphics[scale=0.2]{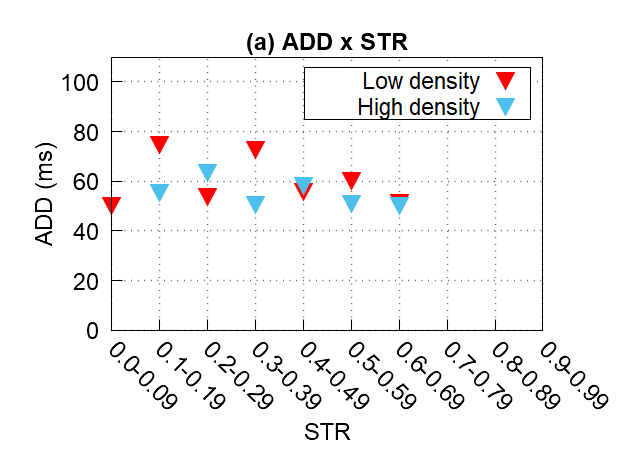}
	\hspace{-0.5cm}
	\includegraphics[scale=0.2]{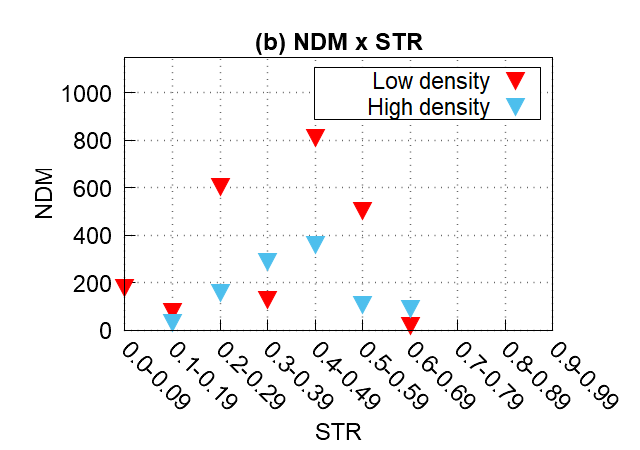}
	\caption{Social influence on the delay and delivery of packets}
    \vspace{-0.3cm}
	\label{fig:cent}
%	\vspace{-0.3cm}
\end{figure}

The graphs in Figure \ref{fig:cent} show the influence of the centrality metrics on the ADD 
%(Graph \ref{fig:cent}(a)) 
and NDM  %(Graph \ref{fig:cent}(b)) 
metrics. The x-axis values of the graphs express the normalized range of STR. In Graph \ref{fig:cent}(a), HD scenario, had a lower ADD because when SOCIABLE selects relay vehicles with higher centrality values (BC, CC, DC and EC), increases the chances of selecting relays better located in the network and close to the BS. While in Graph \ref{fig:cent}(b), LD scenario, SOCIABLE delivered more monitoring packages (higher NDM) since  vehicles spend~more time monitoring the event. Those results show that the social behavior influence the system performance. Despite this, SOCIABLE delivered event data flow under different social~conditions.

\vspace{-0.3cm}

%\as{Graphs on} Figure \ref{fig:str} shows how much the centrality metrics influence on both the ADD (Fig.(a)) and the NDM (Fig.(b)). The x-axis of the graphs express the range of the normalized values of $STR$. The high density scenario has a lower ADD since S-MINUET selects relay nodes with higher centrality values (BC, CC, DC and EC). This increases the chances of selecting relay nodes more well-localized in the network, consequently, closer to the BSs. On the other hand, in the low density scenario, S-MINUET delivers more monitoring packets (higher NMD) since the vehicles spend more time monitoring the event. Such results show that the social behavior affects the system performance. In spite of that, S-MINUET delivers event data under different social (vehicle relationships) conditions.

\begin{comment}

\begin{table}[!ht]
\centering
{
\footnotesize
\begin{tabular}{c|c|c|c|c|}
%\hline
\cline{2-5}
\multirow{2}{*}{} & \multicolumn{2}{c|}{ADD}                          & \multicolumn{2}{c|}{NPME/NPMC}                    \\ 
\cline{2-5}
    & \multicolumn{1}{l|}{LD} & \multicolumn{1}{l|}{HD} & \multicolumn{1}{l|}{LD} & \multicolumn{1}{l|}{HD} \\ 
    \hline
    \multicolumn{1}{|c|}{SOCIABLE} & 56ms & 54ms & 20\% & 3\% \\ 
    \hline
    \multicolumn{1}{|c|}{MINUET} & 60ms & 61ms & 22\% & 4,3\%  \\ 
\hline
\end{tabular}
} %fim tamanho
\end{table}

\end{comment}

\section{Conclusion}
\label{sec:conc}
This work presented SOCIABLE, a system to assist in the dissemination of critical urban events in SIoV environments. It delivers data from critical urban events to external entities by the means of communities created from the social relationships between the vehicles.
Simulations results demonstrated that SOCIABLE selects vehicles that cooperate in the event dissemination and deliver messages to the base station in~real time. Also, it disseminates the data throughout the event duration in a robust fashion. A comparison showed that SOCIABLE transmitted 36.56\% less packets and achieved a maximum delivery delay of 28ms.
%Comparisons made with a solution without social parameters show that SOCIABLE achieves similar or better results with less network overhead.
As future work, we plan to analyze SOCIABLE's performance under events overlapping in time and space and to apply other community mechanisms. 

\section*{Acknowledgment}
We would like to acknowledge the scholarship support of the Brazilian Agencies CAPES and CNPq \#309238/2017-0. 
%This work was supported by CAPES and CNPq Brazilian Agencies: grants \#309238/2017-0. 
\vspace{-0.1cm}
\bibliographystyle{IEEEtran}
%\bibliography{./bibliography/IEEEabrv,./bibliography/IEEEexample}
\bibliography{conference_041818}
\end{document}